\documentclass{article}

\usepackage[preprint, nonatbib]{neurips_2022}

\usepackage[utf8]{inputenc} 
\usepackage[T1]{fontenc}    
\usepackage{hyperref}       
\usepackage{url}            
\usepackage{booktabs}       
\usepackage{amsfonts}       
\usepackage{nicefrac}       
\usepackage{microtype}      
\usepackage{lipsum}
\usepackage{graphicx}
\usepackage{subfigure}
\graphicspath{ {./images/} }
\usepackage{xspace} 
\usepackage{amsmath}
\usepackage{amssymb}
\usepackage{csquotes}
\usepackage[free-standing-units=true]{siunitx}

\begin{document}
\title{A self-supervised scheme for ground roll suppression}
\author{
  Sixiu Liu[1],\and
  Claire Birnie[1],\and
  Andrey Bakulin[2],\and
  Ali Dawood[2],\and
  Ilya Silvestrov[2],\and
  Tariq Alkhalifah[1]\and
  {[1] PSE, King Abdullah University of Science and Technology}\\
  {[2] EXPEC ARC, Saudi Aramco, Dhahran}\\
 {\{sixiu.liu, claire.birnie, tariq.alkhalifah\}@kaust.edu.sa}\\
 {\{andrey.bakulin, ali.dawood.18, ilya.silvestrov\}@aramco.com}\\
}
\date{} 
\maketitle

\begin{abstract}
In recent years, self-supervised procedures have advanced the field of seismic noise attenuation, due to not requiring a massive amount of clean labeled data in the training stage, an unobtainable requirement for seismic data. However, current self-supervised methods usually suppress 
simple noise types, such as random and trace-wise noise, instead of the complicated, aliased ground roll. Here, we propose an adaptation of a self-supervised procedure, namely, blind-fan networks, to remove aliased ground roll within seismic shot gathers without any requirement for clean data. The self-supervised denoising procedure is implemented by designing a noise mask with a predefined direction to avoid the coherency of the ground roll being learned by the network while predicting one pixel’s value. Numerical experiments on synthetic and field seismic data demonstrate that our method can effectively attenuate aliased ground roll.

\end{abstract}


\section{Introduction}

In seismic data, useful signals are often masked by two types of undesired noise: incoherent or coherent. Incoherent, or random noise, is seismic energy that is pixel-wise independent and appears to be random, while those correlated along the time or/and spatial axis are called coherent noise. Among the coherent noise in land seismic data, ground roll is a Rayleigh-type surface wave that is source-generated. Ground roll appears to be dispersive as each wave component with different frequencies propagates at different velocities in near surface, often resulting in seismic data with a “fan” like shape \cite{beresford1988dispersive,deighan1997ground}. With characteristics of low-velocity, low-frequency, high-amplitude, and often spatially-aliased, ground roll is useful for S-wave velocity reconstruction \cite{zhang2021rayleigh}. However, ground roll often masks desired reflected seismic waves in reflection seismology, and its presence can significantly degrade the Signal-to-Noise Ratio (SNR). Thus, ground roll is considered as noise for imaging tasks and should be suppressed in the early stages of seismic processing \cite{claerbout1985ground,saatcilar1988method,henley2003coherent}.

A variety of ground roll attenuation methods have been developed, falling mainly into three categories. The first category is via a specific acquisition geometry, based on the apparent wavelengths of ground roll and body waves \cite{morse1989ground,meunier1998land}. The second category is sparse transform-based methods, which assume linear combinations of a few basis functions in a specific domain can sparsely represent signals, and the ground roll can be removed by a threshold in that domain. Examples include Frequency–Wavenumber (F-K) filtering \cite{foti2002spatial}, radon transform \cite{liu20043,trad2001hybrid,hu2016ground}), wavelet transform \cite{deighan1997ground,chen2017sparsity,lin2022spatial}, curvelet transform \cite{naghizadeh2011ground,liu2018ground,naghizadeh2018ground}, S transform \cite{askari2008ground,tan2013ground,tao2019second}, Shearlet transform \cite{hosseini2015shearlet}, Hilbert transform \cite{jia2020blind}, Karhunen-Loeve transform \cite{liu1999ground,verma2016highly,serdyukov2022ground} and so on. However, the sparsity difference between
signals and ground roll is not always clear in the transformed domain, resulting in an incomplete separation and a certain degree of distortion to the seismic reflections while removing ground roll. The third category is decomposition-based approaches, which assume seismic data can be decomposed into different components, and the principal components can represent signals. These approaches include Singular Value Decomposition (SVD) \cite{kendall2005svd,porsani2010svd,silva2016single}, Empirical Mode Decomposition (EMD) \cite{chen2016ground,xiao2022ground}, Variational Mode Decomposition (VMD) \cite{liu2021ground}, Singular Spectrum Analysis (SSA) \cite{chiu2013coherent,possidonio2021combined}, and Local
Wave Decomposition (LWD) \cite{abbasi2023adaptive}. Other methods include lattice filters \cite{saatcilar1994lattice}, Plane-Wave Destruction (PWD) filters \cite{fomel2002applications}, modelling and inversion approach \cite{strobbia2010surface}, Local Time-Frequency Transform (LTFT) \cite{liu2013seismic}, non-stationary matching filter \cite{jiao2015ground,saatcilar1988method}, polarization analysis \cite{jin2005ground,chen2013robust,wang2017ground}, local band-limited orthogonalization \cite{chen2015ground}, eigenimage filtering\cite{cary2009ground, chiu20193d}, and wavefield continuation \cite{mcmechan1991depth}.

Among the above methods, most of them are effective for ground roll removal when the ground roll is sufficiently different from signals, e.g., in frequency, or the data are regularly and
sufficiently sampled. When the data are insufficiently sampled, or aliased, methods like F-K filtering \cite{foti2002spatial} could fail. A small portion of the above methods, such as Karhunen-Loeve filter \cite{liu1999ground,verma2016highly,serdyukov2022ground}, curvelet  transform \cite{naghizadeh2011ground,liu2018ground}, SVD \cite{kendall2005svd}, polarization analysis \cite{wang2017ground}, Hilbert transform \cite{jia2020blind}, Local Wave Decomposition (LWD) \cite{abbasi2023adaptive}, wavelet transform \cite{deighan1997ground}, Shearlet transform \cite{hosseini2015shearlet}, and eigenimage filtering \cite{chiu20193d}, are targeted to attenuate spatially aliased ground roll. However, when the aliased ground roll and reflections share similar dip or frequency ranges, these methods are also limited in aliased ground roll attenuation.

Recently, Deep Learning (DL) methods, e.g., supervised neural networks, have been widely applied in seismic noise suppression tasks \cite{kaur2020seismic,yang2023deep}. The challenge that traditional supervised methods require training on clean labeled data, which is often unavailable for seismic field records, limits their applications on real data. Therefore, significant effort has been focused on developing deep learning schemes that require no clean training target data. AutoEncoders (AEs) were initially proposed as one such unsupervised approach and have been shown to effectively suppress random noise in land seismic data \cite{chen2019improving,song2020seismic} and coherent, sinusoidal noise in marine seismic data \cite{hamidi2020autoencoder}. However, for large seismic datasets, a small latent space in 
AEs are not able to sufficiently represent seismic data. Recently, two self-supervised denoising training schemes on only noisy data, Noise2Void (N2V) \cite{krull2019noise2void} and Structured Noise2Void (STRUCTN2V) \cite{broaddus2020removing}, have been proposed in biomedical image denoising tasks to 
suppress random and correlated noise, respectively. 
The advent of self-supervised methods in the biomedical field
inspired the self-supervised blind-spot networks, blind-trace networks, and Extended Structured Noise2Void networks, respectively, to suppress seismic random noise
\cite{birnie2021self,birnie2021potential,sun2022seismic,cao2022self}, trace-wise noise \cite{liu2022coherent,liu2022self,romero2022novel,abedi2022multidirectional}, and dipping noise \cite{mosser2022deep}, respectively. It is further found that combining supervised and self-supervised tasks aids the self-supervised suppression of correlated noise in seismic data \cite{liu2022aiding,birnie2022boosting,birnie2022transfer}. Aiming for trace-wise noise removal, the blind-trace networks are classified into input and network masking methods. Input masking methods \cite{liu2022coherent,liu2022self,abedi2022multidirectional} corrupt noisy data through a pre-processing step prior to feeding them as inputs for network training. In contrast, network masking methods \cite{luiken2023integrating,wang2022self} implement deblending by manually modifying the receptive field of a network, following the methodology of \cite{laine2019high}. However, both methods design input or network masks based on the noise characteristics, requiring prior knowledge of the noise. In \cite{birnie2023explainable}, an automatically-designed noise mask obtained via a bias-free network has been utilized to suppress seismic coherent noise successfully. To date, self-supervised blinded networks have shown strong potential on relatively simple noise types, i.e., those with a clear correlation window, out with which the noise is independent. 

DL methods have also shown promising results for suppressing ground roll in seismic field data. Examples include Convolutional Neural Networks (CNNs)-based architectures \cite{zhang2021ground,yang2023deep,cheng2023metaprocessing} and Generative Adversarial Networks (GANs)-based architectures \cite{yuan2020ground,si2020ground,kaur2020seismic}. Due to the lack of clean labeled data, self-supervised learning also developed rapidly into the field of ground roll suppression in  recent years.
Those include CNNs-based architectures \cite{guo2020ground,pham2022physics}, GANs-based architectures \cite{liu2020should,zhao2022unsupervised}, a combination of CNNs and GANs\cite{oliveira2020self}, and Gabor-based Deep Neural Networks (G-DNNs)\cite{liu2023gabor}.
As proof of concept, most of these methods initially focussed on scenarios of either unaliased ground roll, or aliased hyperbolic ground roll \cite{kaur2020seismic,pham2022physics,yang2023deep}. However, in reality, ground roll is typically severely aliased and of fan-shape; with this, removing aliased, fan-shaped ground roll is the common challenge for conventional denoising procedures.

In this study, we adapted the blind-trace scheme aimed for trace-wise noise removal \cite{liu2022coherent} to a blind-fan network to remove aliased, fan-shaped ground roll in seismic data in a self-supervised fashion. In the proposed network, the original noisy data with the ground roll is the network's target, while the 
input is obtained from the same noisy data - a corrupted version in which the blind-fan mask partially conceals the ground roll. The tests on synthetic and field data show that the scheme can effectively suppress ground roll in seismic data. We further investigate the effect of the percentage of corrupted pixels on denoising performance and identify a corruption percentage range that provides an optimal denoising performance.

\section{Theory}
In traditional supervised learning methods, all the pixel values within the network's receptive field contribute to the predicted central pixel.
In N2V \cite{krull2019noise2void}, the central pixel's value is removed from the network's receptive field; in other words, the network cannot use the central pixel's value to predict itself. The removal of the central pixel is obtained by replacing the value of the predicted central pixel, referred to as the active pixel, in the raw, noisy data. The corrupted data and the raw, noisy data are regarded as the network's input and target, respectively. The loss is only computed on the active pixels. Training such a network allows it to predict the value of the active pixel without utilizing the value of the active pixel. With the assumption of pixel-wise independent noise and pixel-wise dependent signals, the signal component of the active pixel can be obtained by the network. However, the N2V scheme fails when the noise is correlated instead of pixel-wise independent.

STRUCTN2V \cite{broaddus2020removing} relaxes the assumption of random noise in N2V and allows us to remove correlated noise by extending the corrupted area from only the active pixel to the area that contains noise that is correlated with that in the active pixel. In the original STRUCTN2V, the cross-correlation of the pure noise decides the corrupted/masked area. For seismic data with trace-wise noise (simply temporally correlated), a vertical blind-trace mask employed on the raw, noisy data corrupts all the pixel values along the residing trace of the selected active pixel \cite{liu2022coherent}. The purpose of the mask is to conceal the noise coherency prior to being fed into the network for training. The corrupted data are, again, regarded as input of the network, while the network's target is the original noisy data. The loss is only computed on corrupted pixels. Training such a network allows it to predict the central pixel's value without utilizing any values from the trace itself. With the assumption of trace-wise noise and pixel-wise dependent signals, the signal component of the predicted pixel can be obtained from the neighboring traces. 

In this study, the fan-shaped ground roll mostly contaminates the seismic data, requiring a blind-fan mask to remove any association between the ground roll in the neighboring area and that in the central pixel. The mask is designed to corrupt the pixel values along the ground roll path of the 
central pixel in the raw, noisy data, in order to 
avoid the network learning the move-out nature of the ground roll. A careful selection of the blind-fan mask structure is required to ensure the noise coherency is concealed while the signals are exposed to the network. A larger noise mask beyond the noise coherency area will result in an incomplete capture of signal information that is necessary for predicting the predicted pixel's value. In contrast, a smaller mask is insufficient to conceal noise information, and the noise outside of the mask can still contribute to the predicted pixel value. 

\begin{figure}[htp]
\centering
\includegraphics[width=1.0\textwidth,height=0.5\textwidth]{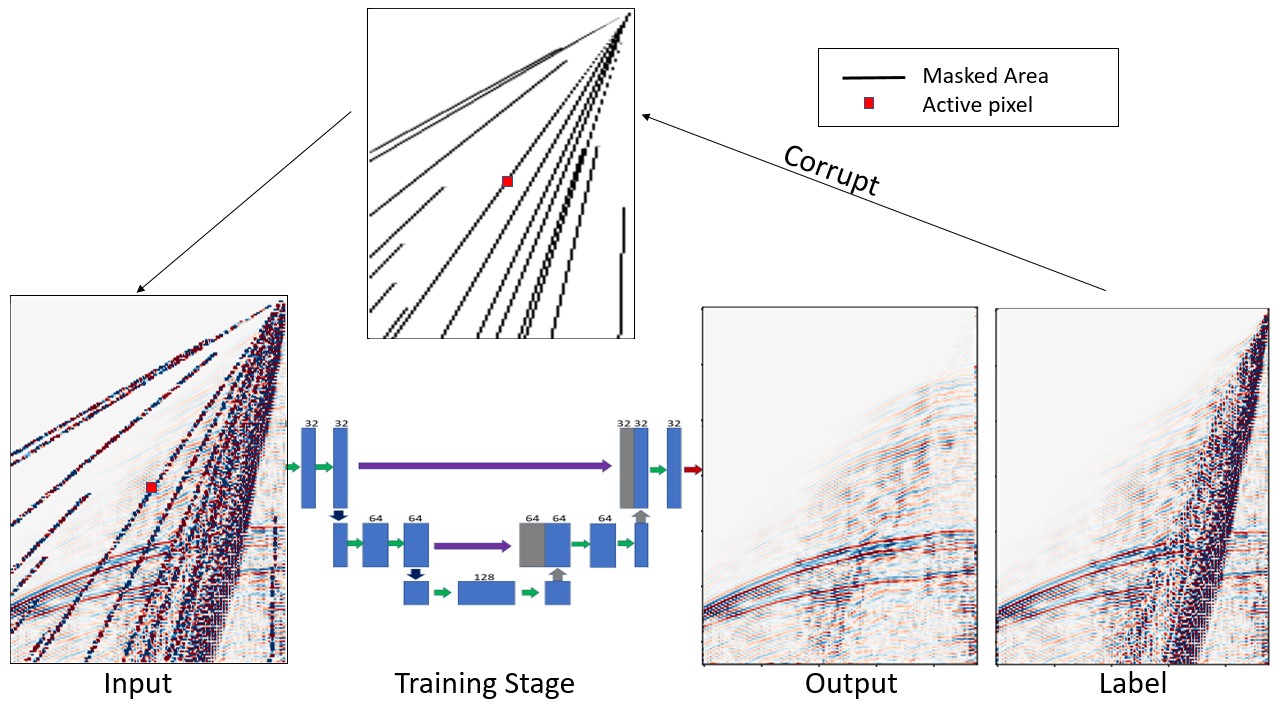}
\caption{Training workflow for ground roll suppression using a blind-fan mask.}
\label{fig:groundroll-scheme}
\end{figure}

Figure \ref{fig:groundroll-scheme} summarizes the proposed training workflow for ground roll suppression using a blind-fan mask. A classical U-Net architecture, including convolutional blocks, each followed by a ReLU activation function and batch normalization, is employed to train the network (styled after \cite{ronneberger2015u}). As shown in Figure \ref{fig:groundroll-scheme}, the label is the noisy data, while the input is a version of the noisy target data corrupted via the blind-fan masking scheme. To corrupt, several active pixels are first randomly selected within the shot gather. Due to the fan shape of the ground roll, which originates from the source location, a line passing through the active pixel is drawn from the source location (assuming the source is at the right corner within the shot gather) to the edge of the shot gather, representing the masking direction. Values randomly chosen from a uniform distribution are used to replace the original values of all the pixels (including the active pixel) along this direction. However, due to the fact that the signal density at earlier times is significantly lower than the later times (See the "Label" in Figure \ref{fig:groundroll-scheme}), an equal number of masking lines (black lines) at each time sample would cause a higher proportion of signals concealment at earlier arrivals than later arrivals. Therefore, removing smaller portions of the masking lines from earlier to later arrivals in the shot gather ensures a similar density of unmasked signals throughout the full recording (See the "Input" in Figure \ref{fig:groundroll-scheme}). The desired output is the denoised data. The loss is shown below:
\begin{equation}
    \mathcal{L} =\sum_{j}\sum_{i}\|\\f\left(\tilde{\boldsymbol{x}}^{ij};\boldsymbol{\theta}\right)-\boldsymbol{x}^{ij}\|_p,
\end{equation}
where $i$ and $j$ stand for the index of the corrupted pixel and the patch.  $\boldsymbol{x}^{ij}$ and $\tilde{\boldsymbol{x}}^{ij}$ are the original pixel value in the target patch and the corrupted pixel value at the same position in the input patch, respectively. The learnt network is $f$, parameterised by $\boldsymbol{\theta}$. The total loss is indicated by $\mathcal{L}$. $\|\cdot\|_p$ represents the $\ell_1$ or $\ell_2$ norm, which can be user-defined. 

Similar to the loss calculation over the corrupted pixels in the blind-trace scheme, the loss here is also measured at the corrupted pixel's locations. The network updates its parameters iteratively through back-propagation.
By training the network on the noisy-corrupted data pairs, one predicted pixel is expected to be constructed without using pixels from the predefined mask direction, instead, using the correlated signals outside of the mask. The best denoising appears when the network completely learns signals but has not started to learn noise yet.

\subsection{Denoising metrics}
Two common metrics, the image Peak Signal-to-Noise Ratio (PSNR) and Mean Square Error (MSE), are adopted to evaluate the denoising performance for synthetic data.

The PSNR (in dB) is computed using the following equation:
\begin{equation}
P S N R=10 \cdot \log _{10}\left(\frac{ {R}^2}{M S E}\right),
\end{equation}
in which
\begin{equation}
M S E=\frac{1}{n}  \sum_{j=1}^{n}[D(j)-d(j)]^2,
\end{equation}
where $n$ and $j$ are the total number and the index of the grid points in the data, respectively. $R$ is the maximum fluctuation in the clean data $D$, and $d$ stands for the data compared to clean data. The higher the PSNR value or the lower the MSE value, the better the denoised data.

\section{Numerical Examples}
\subsection{Synthetic example}
\begin{figure}[htp]
\centering
\includegraphics[width=1.0\textwidth]{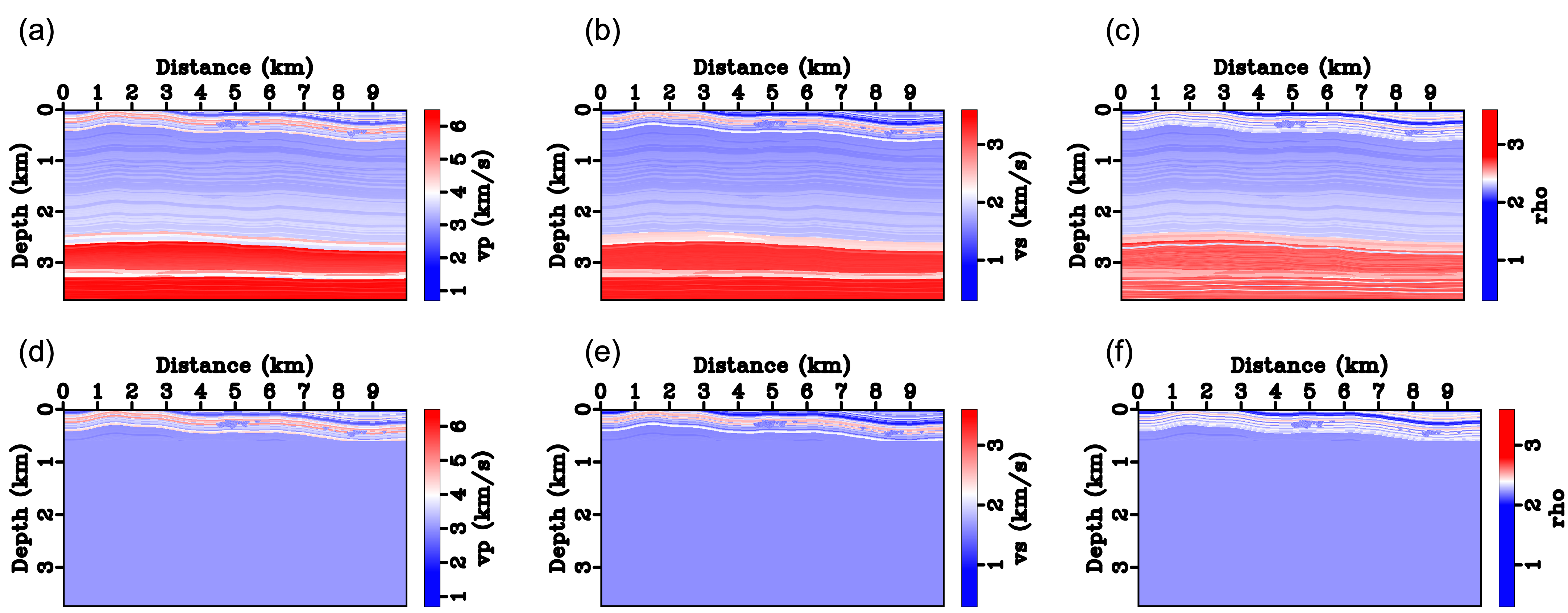}
\caption{Subsurface models from SEAM Arid 2D slice. (a-c): original models; (d-f): the same models as (a-c) but with a homogeneous model below $\sim{600}$ m.}
\label{fig:arid_ss_mods}
\end{figure}

\begin{figure}[htp]
\centering
\includegraphics[width=0.6\textwidth,height=0.5\textwidth]{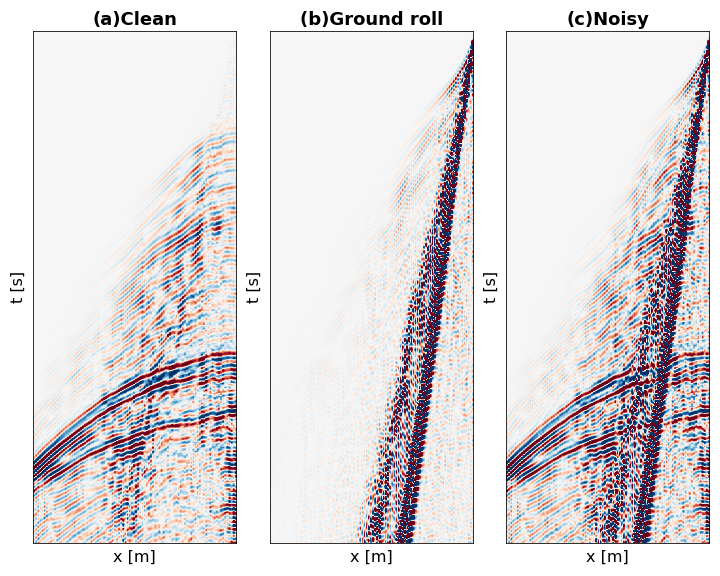}
\caption{(a) reflections, (b) ground roll, and (c) the summed noisy data.}
\label{fig:filterdata}
\end{figure}

To investigate the feasibility of the proposed blind-fan scheme, it is first applied to a realistic synthetic dataset. As illustrated in Figure \ref{fig:arid_ss_mods} (a-c), a 2D section of the SEAM Arid model is used to model the synthetic data. Clean data are acoustically modeled, while the "pure" ground roll is elastically modeled using the same model but a homogeneous half-space below \SI{600}{\metre} (Figure \ref{fig:arid_ss_mods} (d-f)). Considering the general aliasing property of realistic ground roll, both the acoustic data and ground roll are resampled from \SI{2}{\metre} spacing to \SI{50}{\metre} spacing to simulate aliasing. The noisy data (Figure \ref{fig:filterdata} (c)) is generated by summing the clean data and the ground roll, with a weight coefficient on the ground roll to control the noisy data’s SNR. Figure \ref{fig:filterdata} (c) is the target noisy data in the self-supervised denoising scheme. 

For the corruption step, the pixels within the masked areas have their original values replaced by values selected from a uniform distribution over [-0.2, 0.2], similar to the magnitude of the ground roll. The corruption process is repeated per patch per training epoch. Self-supervised schemes do not require a holdout test dataset due to the lack of ground truth labels available. As such, all 53 shot gathers of size $640\times96$ are used for training.  
Based on a hyper-parameter grid search (detailed in Table \ref{tab:para1}) in terms of the maximization of the PSNR and the minimization of the MSE of the denoised image, the optimal training parameters are obtained and shown in Table \ref{tab:para2}. 

\begin{table}[htp]
\centering
\caption{The grid search of hyper-parameters}
\scalebox{0.9}{%
\begin{tabular}{ |p{5cm}|p{4cm}|  }
\hline
   & Parameter options \\
\hline
Batch size & 4,8 \\
UNet depth  & 2,3,4 \\
Loss   & MAE,MSE \\
\hline
\end{tabular}}
\label{tab:para1}
\end{table}

\begin{table}[htp]
\centering
\caption{The optimal training parameters for synthetic and field datasets}
\scalebox{0.9}{%
\begin{tabular}{ |p{5cm}|p{2cm}|p{2cm}|  }
\hline
   & Synthetic &Field \\
\hline
Patch size & 640x96 &400x24 \\
Training size & 53   & 39 \\
Active pixel range  &[5,225] & [5,105] \\
Acceptance range  &[45,95] & [25,45] \\
Noise level to corrupt   &[-0.2, 0.2] & [-0.5, 0.5] \\
UNet depth  & 4 & 4 \\
Batch size & 8 & 4   \\
Initial filters  & 32 & 32 \\
Kernel  & 3 & 3 \\
LR   & 4e-4 & 4e-4 \\
Loss   & MAE & MAE \\
Epochs   & 31 & 5 \\
\hline
\end{tabular}}
\label{tab:para2}
\end{table}

\begin{figure}[!htbp]
\centering 
  \includegraphics[clip,width=1\textwidth,height=0.45\textwidth]{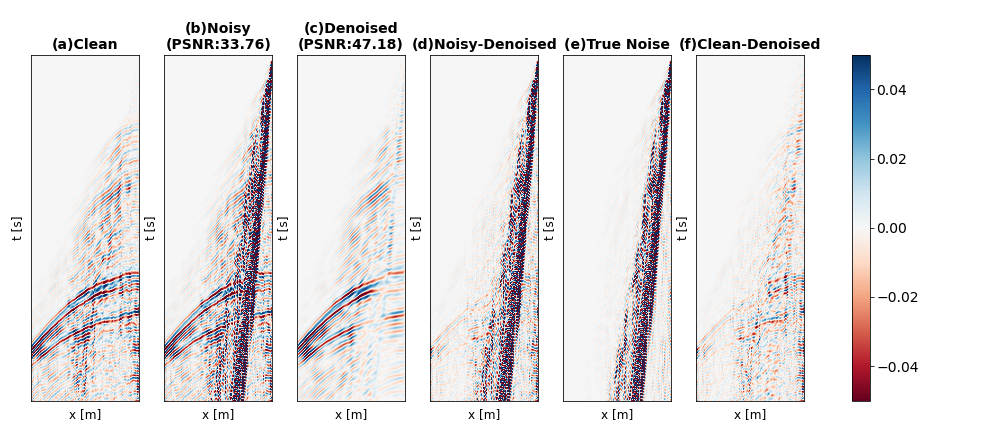}%
  \newline
  \includegraphics[clip,width=1\textwidth,height=0.45\textwidth]{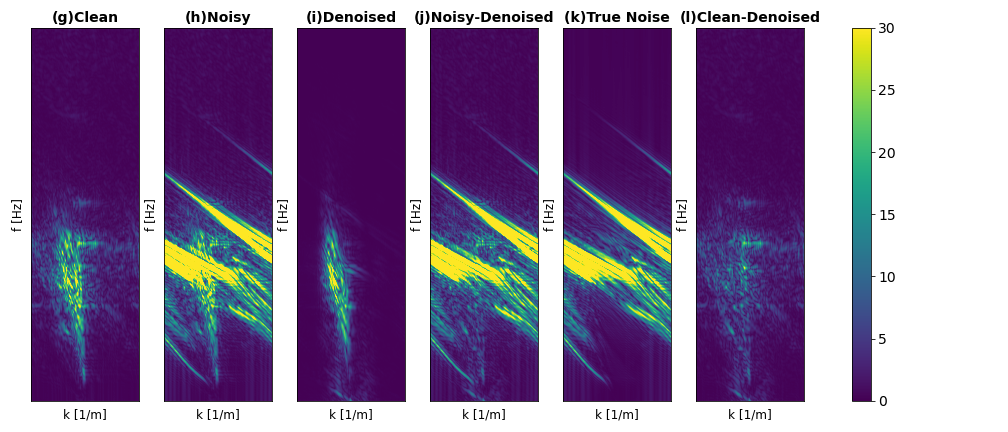}%
\caption{The synthetic results obtained by the trained model. (a)-(f) The clean synthetic, the noisy input,
the denoised data, the removed ground roll, the true noise included in noisy data,
and the difference between the clean and denoised results, respectively. (g)-(l): the corresponding F-K spectra of (a)-(f).}
\label{fig:fixresult}
\end{figure}

Figure \ref{fig:fixresult} illustrates the results of the blind-fan mask scheme on a synthetic seismic dataset.  Figure \ref{fig:fixresult} (a) is the noise-free data for comparison. Figure \ref{fig:fixresult} (b) shows the high-amplitude, spatially aliased ground roll that masks near-offsets
and far-offsets reflections from early to later times, resulting in a PSNR of 33.76. As shown in Figures \ref{fig:fixresult} (c), the proposed procedure effectively removes ground roll and recovers the previously concealed reflection information, resulting in a much higher PSNR of 47.18. Compared to the true ground roll (Figures \ref{fig:fixresult} (e)) added in Figure \ref{fig:fixresult} (b), the removed ground roll (Figures \ref{fig:fixresult} (d)) indicates the majority of the ground roll has been suppressed by the procedure; however, part of the later seismic arrivals have also been simultaneously removed. Figure \ref{fig:fixresult} (f) indicates the denoising scheme could also cause some amount of signal damage, especially at the ground roll contaminated area. Compared to the weak energy of the pure reflections in the F-K domain in Figure \ref{fig:fixresult} (g), Figure \ref{fig:fixresult} (h) shows how the aliased ground roll strongly dominates the noisy data. While Figure \ref{fig:fixresult} (i) rarely suffers from the aliased ground roll, Figure \ref{fig:fixresult} (j), (k), and (l), again, demonstrate the powerful ability of the procedure in removing the aliased ground roll. 

\subsubsection{Effect of number of active pixels}
In all previous blinded networks, either N2V \cite{krull2019noise2void} and STRUCTN2V \cite{broaddus2020removing} from the biomedical field, or blind-spot \cite{birnie2021potential} and blind-trace networks \cite{liu2022coherent} from seismic field, the number of active pixels chosen in a patch is often optimally tuned. The number of active pixels can be regarded as an extra constraint to the training loss in the network training stage, resulting in a trade-off between the training epoch number and the active pixel number. A larger number of active pixels results in a larger corruption percentage and more pixels contributing to the loss at one epoch, requiring fewer training epochs to learn to denoise. However, this may lead to too strict noise concealment and sometimes signals concealment simultaneously. In contrast, a small number of active pixels results in a smaller corruption percentage and fewer pixels contributing to the loss at one epoch, requiring more training epochs. This may lead to too much remaining noise exposure to the network. The best denoising performance is always found in the scenario with an optimal number of active pixels, with an optimal percentage of corruption.

\begin{figure}[htp]
    \centering
    \includegraphics[width=0.95\textwidth,height=0.55\textwidth]{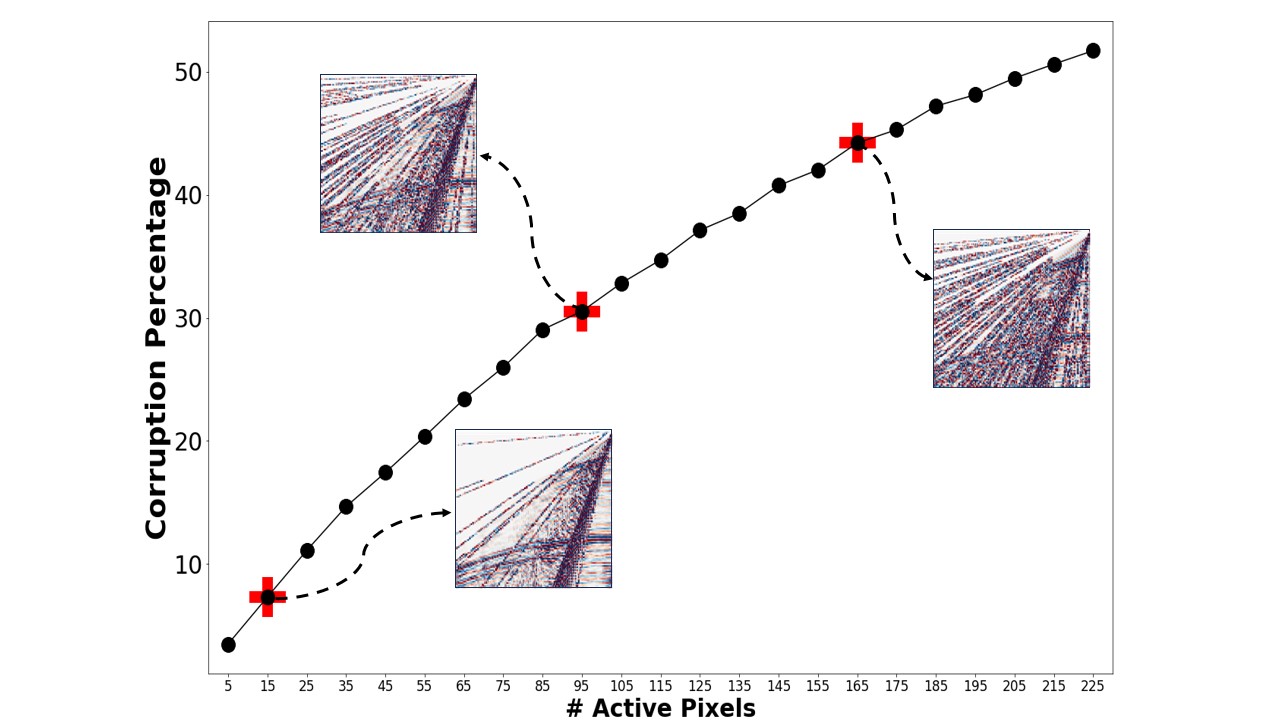}
    \caption{Corruption percentage changes over active pixel numbers. The three red crosses illustrate the corresponding corrupted data, for a given active pixel value.}
    \label{fig:corr}
\end{figure}

Similar to the previous studies, we analyze the effect of the number of active pixels on denoising performance on synthetic data. A shot gather includes 61440 (640x96) pixels, among which we select some as active pixels. Given a certain active pixel number, the corruption percentage is the ratio of the number of corrupted pixels obtained by the blind-fan mask and all the pixel numbers within a shot gather. Averaging over all the shot gathers, we illustrate that the corruption percentage is almost linearly increasing with the active pixel number, with a larger active pixel number representing a larger corruption percentage, as shown in Figure \ref{fig:corr}. Therefore, the number of active pixels can be directly used to analyze the effect of the corruption percentage on denoising performance.

\begin{figure}[htp]
    \centering
    \includegraphics[width=1\textwidth,height=0.55\textwidth]{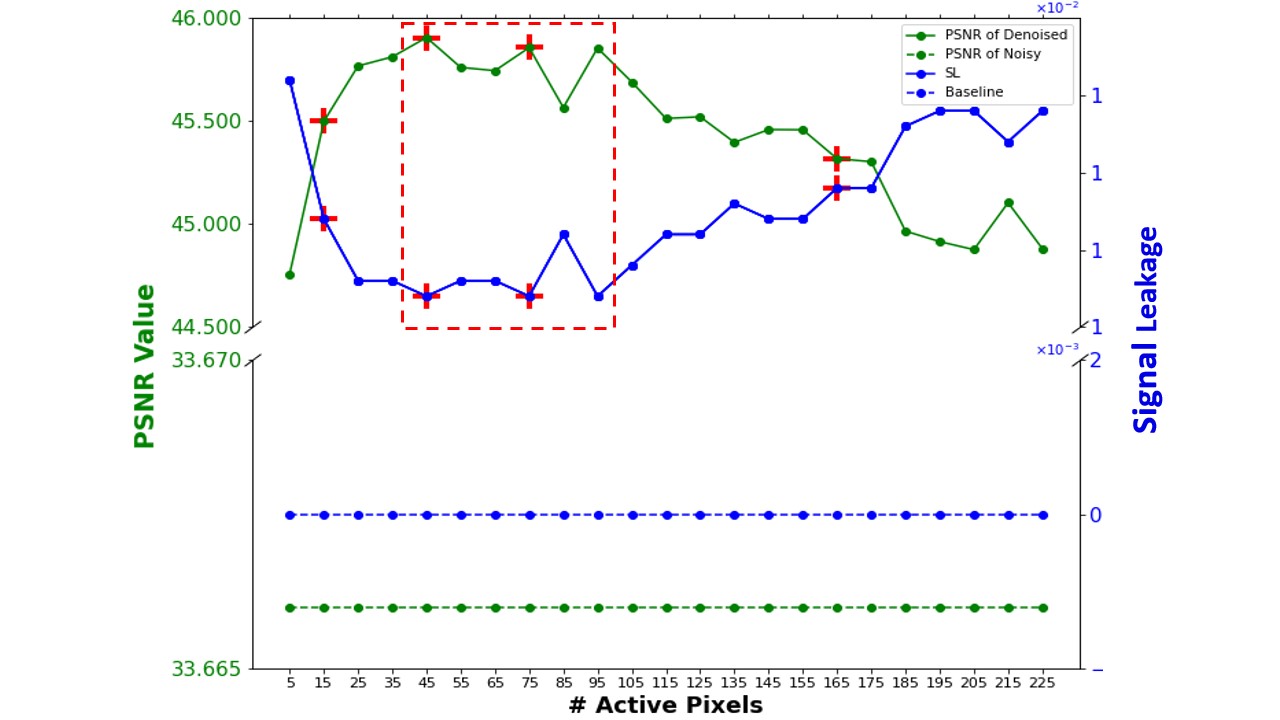}
    \caption{PSNR and signal leakage value changes over active pixel numbers. The red dashed box illustrates the acceptance range, and the red cross illustrates the four active pixel samples for showing the denoising performance.}
    \label{fig:sa}
\end{figure}

Figure \ref{fig:sa} indicates the effect of the number of active pixels (or corruption percentage) on denoising performance. The MSE value evaluating the difference between the clean and denoised data indicates the volume of the signal leakage. The metrics of PSNR and signal leakage are compared in Figure \ref{fig:sa}. The noisy data have an averaged PSNR value, which is slightly lower than 33.67. With a much higher PSNR value than the noisy data for all active pixel numbers, the PSNR of denoised data first increases before the range (shown by the red dashed box): the range between 45-95, and starts to decrease after this range. This range, referred to as an acceptance range, indicates an active pixel range where the denoising metrics and denoising performance are comparably good. The signal leakage shows an opposite trend. However, both curves indicate the same denoising behavior: the network performs almost equally well for some active pixel numbers that are in the middle of the parameter settings, which can be explained as follows. The purpose of the blind-fan mask is to corrupt the ground roll information by random values to hide the noise coherency while keeping the signal components exposed to a predicted pixel. For a number of active pixels prior to the acceptance range (or smaller than 45), some ground roll components within the receptive field are not fully masked; therefore, ground roll can still contribute to a predicted pixel. For a number of active pixels after the acceptance range (or larger than 95), the signal components are also masked simultaneously along with the ground roll concealment, resulting in a loss of signals that are required to recreate the predicted pixel. The existence of this acceptance range for synthetic data guides the application of field data. 

\begin{figure}[htp]
    \centering
    \includegraphics[width=1.0\textwidth,height=0.87\textwidth]{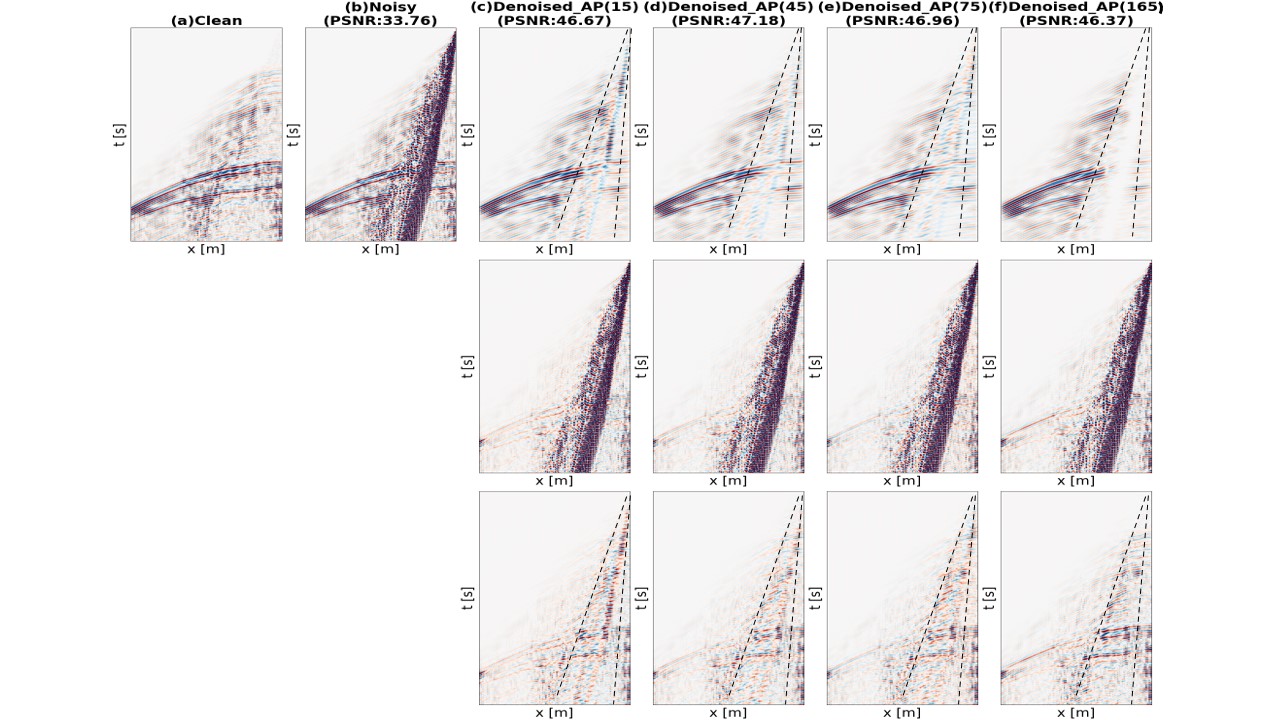}
    \caption{The synthetic results obtained by the trained model, trained with different active pixel numbers. (a)-(f) The clean synthetic, the noisy input, the denoised result trained with active pixel numbers of 15, 45, 75, and 165, respectively. Top: clean, noisy, and denoised data; Middle: removed ground roll; Bottom: signal leakage. The black dashed lines illustrate the interest area.}
    \label{fig:diff-ap}
\end{figure}

\begin{figure}[htp]
    \centering
    \includegraphics[width=1\textwidth,height=0.87\textwidth]{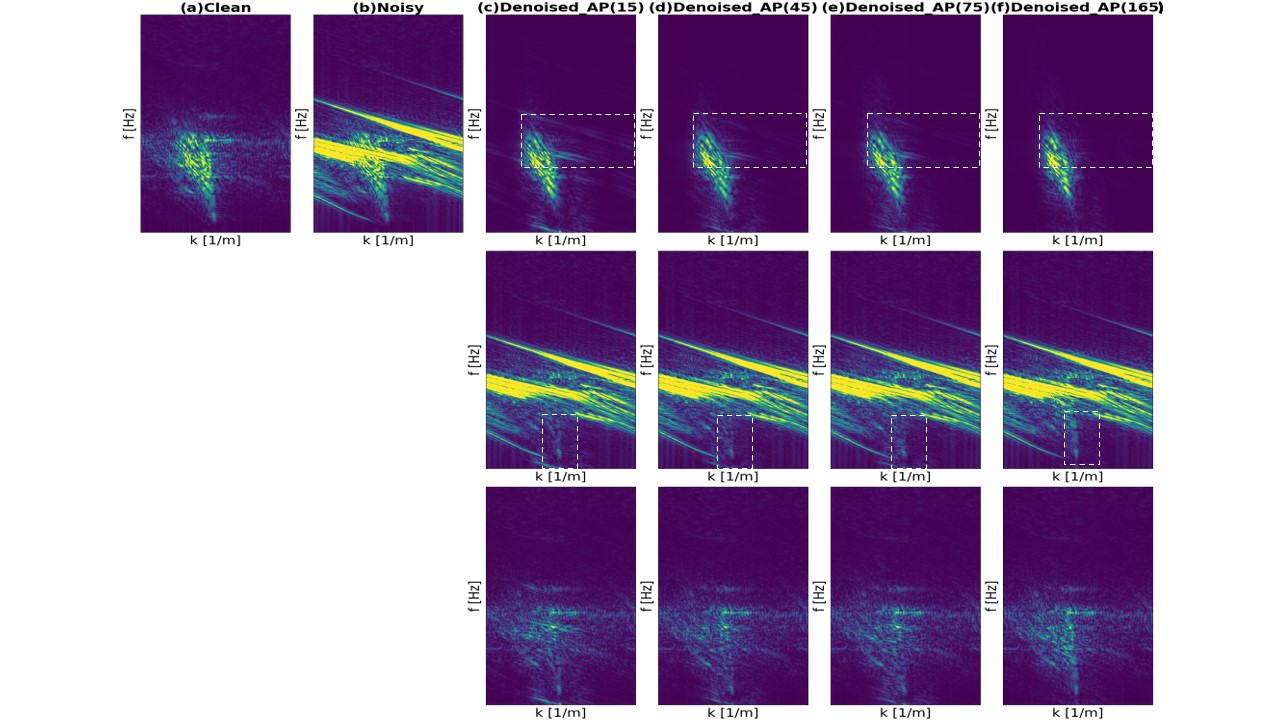}
    \caption{The corresponding F-K spectra of Figure \ref{fig:diff-ap}. Top: clean, noisy, and denoised data; Middle: removed ground roll; Bottom: signal leakage. The white dashed boxes illustrate the interest area.}
    \label{fig:diff-ap-spe}
\end{figure}

As shown in Figure \ref{fig:sa}, the metric values in the acceptance range are close to each other. Two samples (active pixels of 45 and 75, annotated by the red cross) within the acceptance range and two (active pixels of 15 and 165, annotated by the red cross) out of the acceptance range are chosen to compare the denoising performance. In Figure \ref{fig:diff-ap}, four denoising performances with different active pixels: 15, 45, 75,  and 165, are compared. Highlighted by black dashed lines in Figure \ref{fig:diff-ap}, it is observed that the active pixel of 45 and 75 (Figure \ref{fig:diff-ap}d and \ref{fig:diff-ap}e) can both better suppress ground roll in terms of noise removal and signal leakage. For active pixels of 15 and 165 (Figure \ref{fig:diff-ap}c and \ref{fig:diff-ap}f), the denoised data struggles with either too much remaining ground roll or too much loss of signals, especially at the ground roll contaminated area. In Figure \ref{fig:diff-ap-spe}, the F-K spectra of Figure \ref{fig:diff-ap} are compared. Highlighted by white dashed boxes in Figure \ref{fig:diff-ap-spe}, the active pixels of 45 and 75 (Figure \ref{fig:diff-ap-spe}d and \ref{fig:diff-ap-spe}e) have comparably better results than the other two. The ones with active pixels of 15 and 165 (Figure \ref{fig:diff-ap-spe}c and \ref{fig:diff-ap-spe}f) struggle with either incomplete suppression of aliased ground roll or too many signals removed, respectively.

\subsection{Field data application}
The field data are shot gathers \cite{zhang2021rayleigh} collected at the Mirandola site in Italy. The shot gathers are of 400 time samples with intervals of 0.25 ms and 24 receivers with 1 m sampling, providing 39 shot gathers of $400\times24$. The ground roll energy in the original shot gathers is much higher than the reflections, especially at early arrivals and near offsets. To avoid the network's learning target bias (bias to either ground roll or reflections), the energy between the ground roll and the reflections in the original shot gathers is first balanced through a trace-wise normalization procedure. 

For field data, the pixels within the masked areas have their original values replaced by values selected from a uniform distribution over [-0.5, 0.5], given the field data is normalized to 1. The corrupted data, or the input, is modified per patch and iteration. Again, all the available 39 shot gathers of size $400\times24$ can be used to train the network. Table \ref{tab:para2} shows the optimal training parameters selected via trial-and-error.

\begin{figure}[htp]
    \centering
    \includegraphics[width=1\textwidth,height=0.65\textwidth]{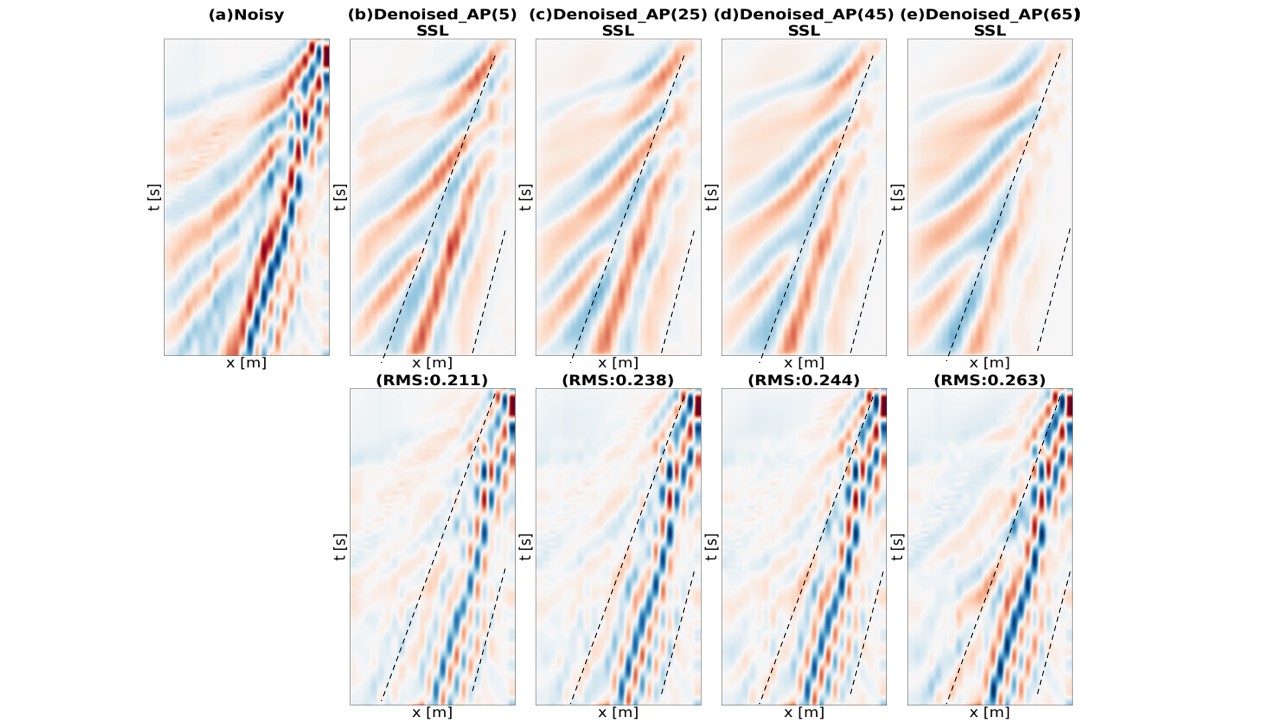}
    \caption{The field results obtained by the trained model, trained with different active pixel numbers. (a)-(e) The noisy input, the denoised result trained with active pixel numbers of 5, 25, 45, and 65, respectively. Top: noisy and denoised data; Bottom: removed ground roll. The black dashed lines illustrate the interest area.}
    \label{fig:field_ap5-65}
\end{figure}

Similarly, the analysis of the number of active pixels on denoising performance is applied to field data. A shot gather includes 9600 ($400\times24$) pixels, among which we select some as active pixels, which vary from 5 to 105, with a step of 10. The acceptance range (again, the active pixel range for best denoising performance) for field data is found to be the range between 25-45, based on the trade-off between the volume of the remaining signals and the removed ground roll. Figure \ref{fig:field_ap5-65} illustrates the results of the blind-fan mask scheme on the field seismic dataset, with different numbers of active pixels, e.g., 5, 25, 45, and 65. 
It is observed that the active pixel of 25 (Figure \ref{fig:field_ap5-65}c) and 45 (Figure \ref{fig:field_ap5-65}d) show comparably better denoising performance than the other two, in terms of the volume of the removed ground roll and signal leakage. However, the active pixel of 5 (Figure \ref{fig:field_ap5-65}b) struggles with too much remaining ground roll, and the active pixel of 65 (Figure \ref{fig:field_ap5-65}e) struggles with too much signal loss. 

\begin{figure}[htp]
    \centering
    \includegraphics[width=1\textwidth,height=0.6\textwidth]{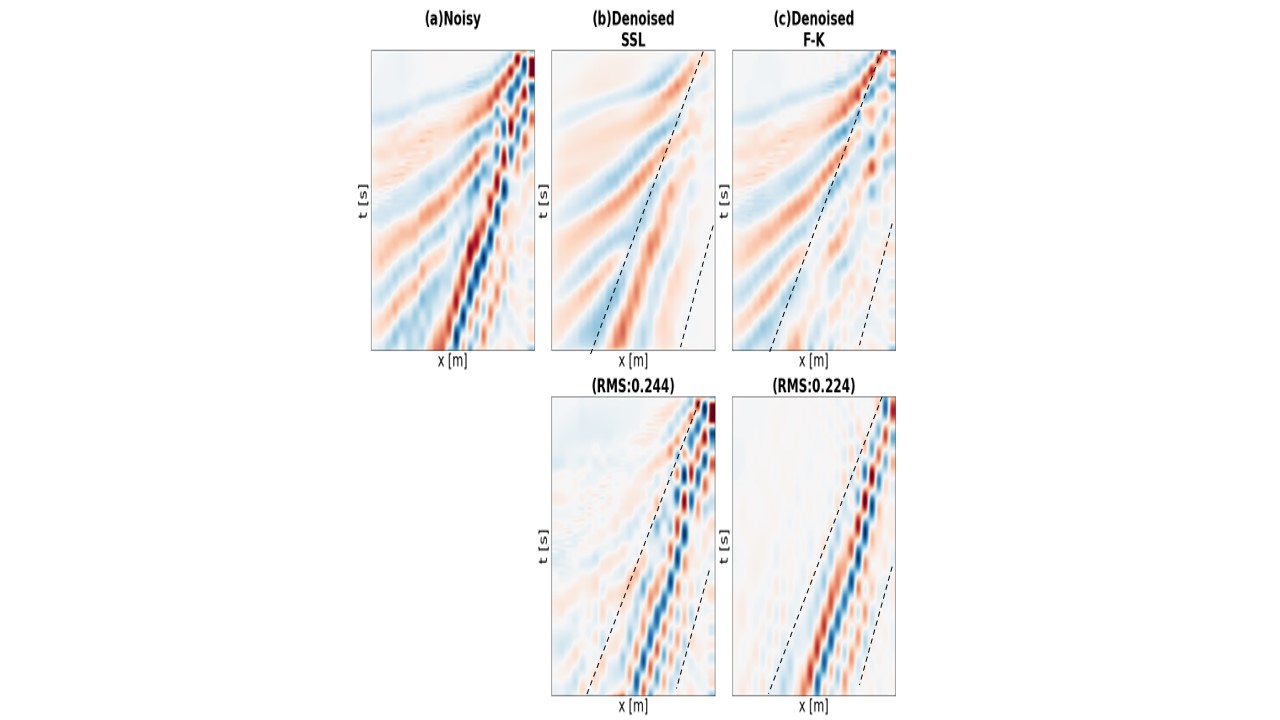}
    \caption{The field results obtained by the trained model, with active pixel number, 45. (a)-(c) The raw field data, the denoised result by SSL, and the denoised result by F-K. Top: noisy and denoised data; Bottom: removed ground roll. The black dashed lines illustrate the ground roll contaminated area.}
    \label{fig:field}
\end{figure}

\begin{figure}[htp]
    \centering
    \includegraphics[width=1\textwidth,height=0.6\textwidth]{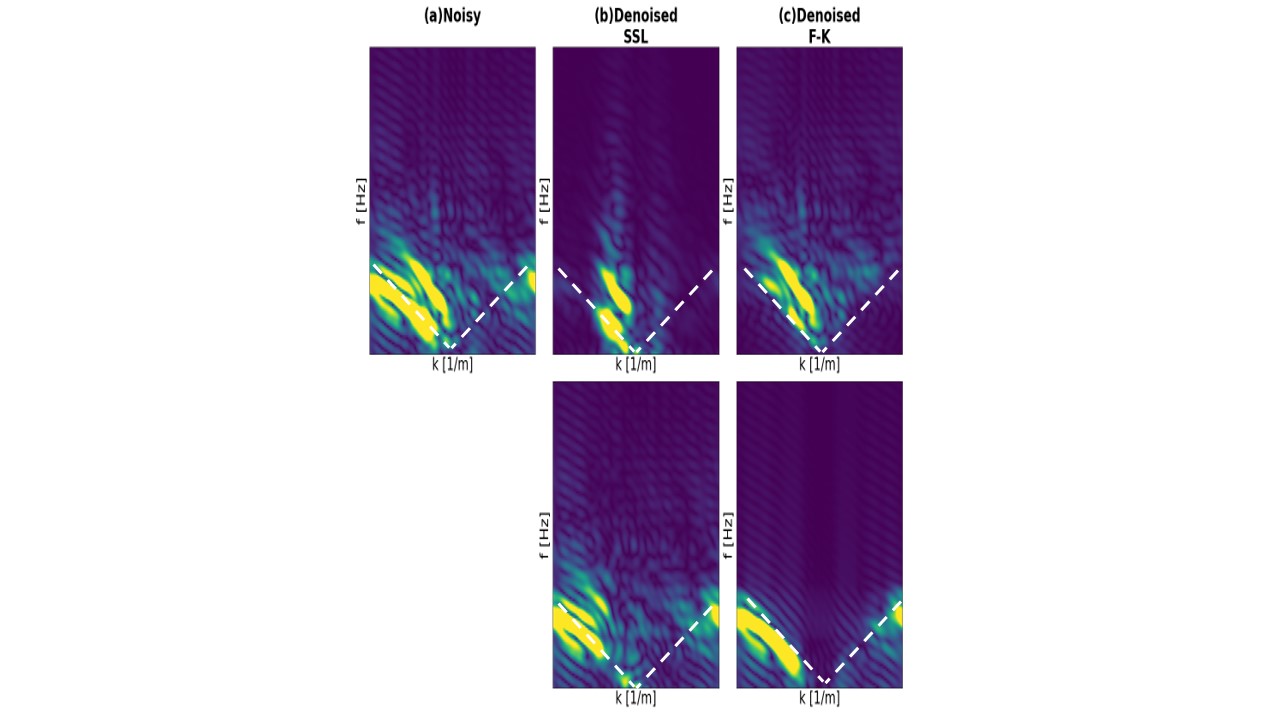}
    \caption{The corresponding F-K spectra of Figure \ref{fig:field}. Top: noisy and denoised data; Bottom: removed ground roll. The white dashed lines illustrate the separation between ground roll and reflections.}
    \label{fig:field-spe}
\end{figure}

With the acceptance range for field data being 25-45, we illustrate the results of the blind-fan mask scheme on the field seismic dataset, with one of the optimal numbers of active pixels, 45. A conventional method: the F-K filter, is applied to field data to be compared with the self-supervised learning (SSL) method. As seen in the noisy data in Figure \ref{fig:field}(a), the reflections are severely masked by the ground roll, especially at near offsets and earlier times. As shown in Figure \ref{fig:field} (b) and (c), while both methods can retain most of the reflection energy, the SSL scheme removes more ground roll at early times and less ground roll at later times than the F-K filter. The difference between the original noisy data and denoised results shows a significant removal of ground roll, with slightly more signal leakage of the SSL scheme than the F-K filter. The black dashed lines in Figure \ref{fig:field} show the ground roll contaminated area. In Figure \ref{fig:field-spe}, the F-K spectra of Figure \ref{fig:field} are compared. Highlighted by white dashed lines in Figure \ref{fig:field-spe} (a)-(c), the separation in F-K spectra by SSL is not complete, while the one by the F-K filter is relatively complete.

\section{Discussion}
We have seen how the self-supervised bind-fan networks are utilized to suppress ground roll in aliased seismic data. Initially targeted for trace-wise noise removal, the blind-trace mask is converted to a bind-fan mask for successful ground roll attenuation in field data, resulting in a comparable denoising performance to the conventional F-K filter. 
While an F-K filter is costly in finding an accurate cut-off velocity to separate signals and ground roll, the proposed blind-fan scheme requires a careful selection of the mask design. The successful suppression of ground roll by the proposed scheme indicates the potential for other coherent seismic noise attenuation, such as hyperbolic ground roll or interference noise in marine seismic data. With this, the following key aspects still need to be discussed.

\subsection{Loss convergence}
For self-supervised blinded networks, with noisy data being the target, the loss converging to zero is not desired. For example, given a perfectly designed mask, the proposed blind-fan network could be trained to infinity to reconstruct the signal components while totally ignoring the ground roll components from the source location, as it should never recreate the noise. However, a perfect mask implementation is hard, which will cause a slight noise leakage from the unmasked area to a predicted pixel with long network training. Longer training after the optimal point would cause the identity recreation of both signals and noise, as the target is the noisy data. Therefore, an optimal denoising loss is where the signals are thoroughly learned, but the noise is not. 

\subsection{Network robustness in inference stage}
\begin{figure}[htp]
    \centering
    \includegraphics[width=0.9\textwidth,height=1.2\textwidth]{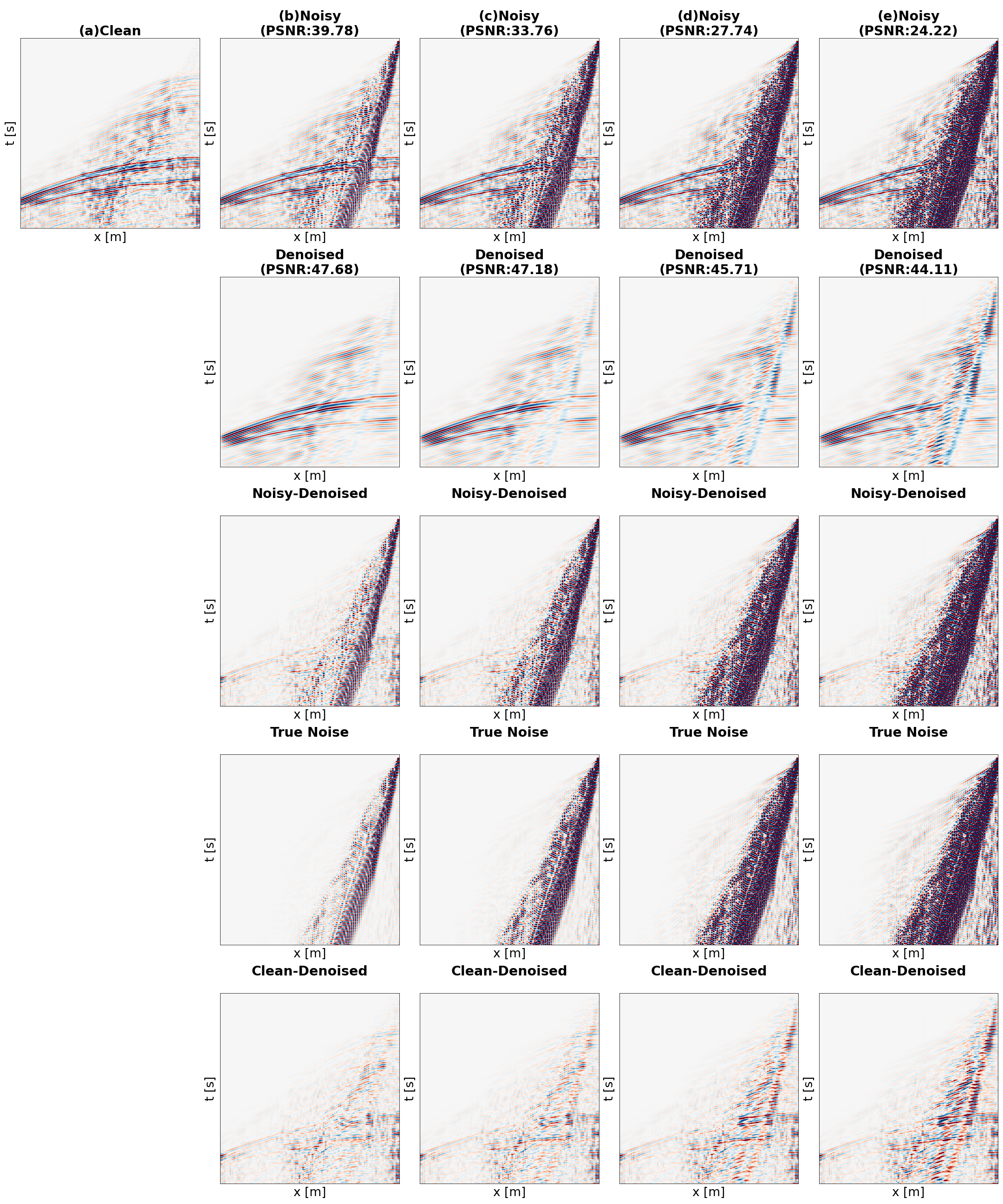}
    \caption{The synthetic results obtained by the trained model, trained on a certain noise level but applied to synthetic datasets with different levels of ground roll. (a)-(f) The clean synthetic, the trained model applied to synthetic datasets with noise levels lower, same, higher, and much higher than the trained noise level, respectively. 1st panel: clean and four noisy data; 2nd panel: denoised data; 3rd panel: removed ground roll; 4th panel: true noise; 5th panel: signal leakage.}    
    \label{fig:adap}
\end{figure}

A common topic in deep learning denoising tasks in terms of network robustness is how a network trained on a certain noise level performs on seismic data that have noise levels that are not seen in the training stage. Therefore, we applied the network trained on a certain noise level, with a PSNR of 33.76 (Figure \ref{fig:adap}(c)), to seismic data with the same clean contents but with three different noise levels, with PSNR of 39.78, 27.74, and 24.22, respectively, as shown in Figure \ref{fig:adap} (b), (d), and (e). It is observed in Figure \ref{fig:adap} (b)-(e) that the trained network can obtain much cleaner data than the noisy data, with a much higher PSNR increase and significantly comparable removed ground roll with the true noise. This indicates that a single-trained blind-fan network can be reasonably applied to suppress noise levels that are much lower (Figure \ref{fig:adap}(b)) or higher (Figure \ref{fig:adap}(d) and (e)) in inference stage than the noise level that is seen in the network training.

\subsection{Limitations}
As mentioned in the theory section, the first step of the masking is to select several active pixels to decide the ground roll direction from the source location; therefore, the proposed method is suitable for denoising the full shot gathers that include the source location. The scheme for denoising some sections of the shot gathers needs to be further investigated.

\section{Conclusions}
Blind-trace and blind-spot schemes have shown to be successful suppressors of trace-wise noise and random noise in seismic data, without requiring clean labels. In this study, we proposed a blind-fan mask to suppress ground roll in a self-supervised fashion, by using the prior knowledge that the ground roll is source-generated. By masking the noise coherency along the noise direction in the original, raw, noisy data, using a fan-shaped mask, a corrupted version of noisy data is generated to be the network's input. The masking scheme blinds the ground roll information and only exposes the signal information within the receptive field to be learned for a central pixel’s prediction. Training a network with training pairs of target noisy and corrupted data allows
a network to predict a central pixel's value by only utilizing the pixels outside the masked area. The ground roll suppression results on both synthetic and field data show the effectiveness of the proposed blind-fan network in removing aliased ground roll in seismic data, with a slight amount of signal leakage. Furthermore, we show how a batch of different active pixel numbers within the acceptance range provides comparable good denoising results and how a single-trained network can be applied to attenuate noise levels not seen in training. Moreover, we show how the 
self-supervised scheme performs similarly to a conventional denoising method, the F-K filter, on field data. 

\section{Acknowledgment}
The authors thank King Abdullah University of Science  Technology (KAUST) Seismic Wave Analysis Group and Saudi Aramco for supporting this research and also, in particular, Dr. Yuanyuan Li, Dr. Nick Luiken, and Dr. Zhendong Zhang for insightful discussions. For computer time, this research used the resources of the Supercomputing Laboratory at KAUST in Thuwal, Saudi Arabia. 
\newpage
\bibliographystyle{plain}  
\bibliography{reference}
\end{document}